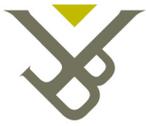

**N. Rons, L. Amez**

N. Rons is Coordinator of Research Evaluations and Policy Studies, Research Coordination Division, Research & Development Department, Vrije Universiteit Brussel (VUB), B-1050 Brussels, Belgium; E-mail: nrons@vub.ac.be.

L. Amez is Policy Collaborator for Research Publications, Research Coordination Division, Research & Development Department, Vrije Universiteit Brussel (VUB), B-1050 Brussels, Belgium; E-mail: Lucy.Amez@vub.ac.be.

# Impact Vitality: An indicator based on citing publications in search of excellent scientists

## Abstract

This paper contributes to the quest for an operational definition of 'research excellence' and proposes a translation of the excellence concept into a bibliometric indicator. Starting from a textual analysis of funding program calls aimed at individual researchers and from the challenges for an indicator at this level in particular, a new type of indicator is proposed. The Impact Vitality indicator [RONS & AMEZ, 2008] reflects the vitality of the impact of a researcher's publication output, based on the change in volume over time of the citing publications. The introduced metric is shown to posses attractive operational characteristics and meets a number of criteria which are desirable when comparing individual researchers. The validity of one of the possible indicator variants is tested using a small dataset of applicants for a senior full time Research Fellowship. Options for further research involve testing various indicator variants on larger samples linked to different kinds of evaluations.

## Introduction

Excellence plays an increasingly important role in research policy. Countries worldwide have introduced a variety of excellence programs, aiming to steer substantial amounts of funding towards the best or most promising researchers or research centres. The European Commission introduced Networks of Excellence as a part of its Sixth Framework Program (2002-2006) to strengthen excellence in particular research topics. More recently, the European Research Council awarded its first Advanced Investigator Grants for exceptional research leaders. Excellence grants are generally awarded based on a qualitative assessment by an expert committee. While this practice is widely accepted, the process of peer review is also subject to criticism. Weaknesses observed in a context of decisions for funding include policy issues, bias and limited selective power [CHUBIN AND HACKETT, 1990; LANGFELDT, 2004] and the influence of chance, scales and budget constraints [COLE ET AL., 1981; LANGFELDT, 2001]. Peer assessments are therefore often assisted, informally or formally, by 'more objective' quantitative measures [MOED, 2005].

The concept of excellence is complex and multidimensional in nature. Qualities looked for in the individual scholar involve the talent for creating innovative knowledge, successful dissemination and peer recognition, as well as managerial capacities to guide other scholars and to assure financial support. Excellence is also a comparative concept, pointing to the ability to surpass others in quality, to be the best within a chosen benchmark set [TIJSSEN, 2003]. The major question is how to recognize an excellent scientist among a selection of very good researchers, helped by adequate measures. Obviously, different measures are needed to represent different kinds of research outputs (such as obtained funding, awards, etc.). This paper concentrates on publication activity as a prominent aspect in the assessment of a scholar's performance in any domain. It is indeed in the nature of good scientific practice to communicate one's knowledge and findings to peers and scientific publication is probably the only activity that is sufficiently uniform to be compared in an internationally oriented selection process. Bibliometric indicators describing the publication activity and its impact have been widely applied at higher aggregation levels, such as research groups, departments, universities or countries. At the level of individuals however, most of these measures are much less appropriate and there is a clear need for new, dedicated indicators to help select the best scientists that will be supported in their leading role.

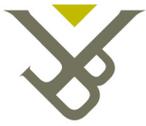

## Methodology

Several bibliometric tools proposed to identify excellence focus on high-end performances for generally established bibliometric indicators (i.e. not specific for assessments at excellence level). A crown indicator value *CPP/FCSm* above 3 generally indicates that a group can be considered excellent [VAN RAAN, 2000]. Other tools focus on 'top' segments such as the most highly cited papers, publications in the most prestigious journals or citations from high impact journals [TIJSSEN ET AL., 2002; VAN LEEUWEN ET AL., 2003]. When concentrating on assessments of individual researchers at excellence level, two types of problems are encountered. Firstly, known caveats are amplified when bibliometric indicators are applied to smaller samples of publications, where the way one publication is taken into account may considerably influence results. Secondly, while standard bibliometric indicators can give a good indication of subjects with very good performance, they are not necessarily adequate to distinguish excellent subjects from the very good ones, because excellence may to a large extent involve other capacities than the ones they measure.

The approach used in this paper starts from a description of what such an indicator should measure and from the technical characteristics it should obey. A first part examines the selection requirements of funding programs to screen which distinguishing capacities are associated with excellent researchers. The aim is to come to a workable definition of excellence representing real selection practice. The second part examines what is expected from a bibliometric indicator applied at individual level. It lists a number of technical requirements and challenges that the indicator to be constructed needs to respond to. Joining both the conceptual and the structural track, a new indicator is introduced aiming to identify individual excellence based on the vitality of the impact of a researcher's publication output. Finally, the paper discusses how this indicator meets various requirements, illustrated with results for a dataset of applicants for senior full time Research Fellowships.

## Towards an operational definition of excellence

While excellence constitutes the primary aim of dedicated funding programs, the concept is rarely well defined. Program calls display a variety of vague phrasings, such as 'internationally recognized', 'excellent research records' and 'internationally accepted criteria for excellence'. More operational appreciations associated with excellence, selected from a number of funding programs are: at the forefront / substantial contribution to the development of the field' (Methusalem program, Belgium); prominence in the field (Odysseus program, Belgium); continued performance at top level for a considerable time (Spinoza prize, The Netherlands); increasing research productivity (Leibniz prize, Germany); lead author of papers that have made a significant impact in their field, i.e. highly cited (Research professor recruitment award, Ireland). Tangible criteria for excellence are put forward only in a limited number of cases. The threshold for excellence is left to the judgment of the reviewers, based on their knowledge of the domain. The paths of reflection that lead to their separation of excellent researchers from the others are mostly unknown. Yet a clear idea of what excellence means in practical terms is crucial for the construction of adequate indicators. Building on the objectives of the programs and focusing on what they imply when it comes to judging publication activities, the following definition of excellence is proposed:

*An excellent researcher is prominently present in the field, continuously publishing new knowledge and ideas over a long period of time. As an established reference in the field, his/her contributions are eagerly followed by colleagues and his/her ideas are picked up fast in their further research. As such, he or she is a central figure in a strong research dynamic, at the level of the researcher's own research team as well as for the research area as a whole, increasing both volume and impact of research in the field.*

In this paper, an indicator is proposed which optimally respects this definition. The activity of publication that it centers on refers to written research output, such as represented in international databases, and the notion of impact is narrowed down to citation impact present in these databases. International publication and citation databases holding a large and representative part of the most important scientific outputs in many fields, offer a good general basis for constructing indicators in the context of scientific communication. They cannot be a good basis however for constructing indicators for publication cultures that primarily use media outside their scope, possibly in past or future research traditions, or in currently insufficiently covered disciplines.

### Indicator requirements and challenges

Bibliometric indicators are not regarded as substitutes for peer review and the limitations of their potential use have been widely studied. On the other hand, as publication is the most important way to disseminate scientific knowledge, citations are accepted to hold crucial information about scientific impact and influence for a wide range of disciplines. At the level of individual scientists however, the effects of bias and other shortcomings are amplified and it becomes of outmost importance to both control the constituent factors of the indicators and spell out in detail which technical requirements the metric has to obey. Table 1 lists a number of requirements and challenges that indicators applied to individual scientists should be able to deal with. Each item is discussed briefly below.

Table 1. Requirements and challenges for indicators applied to individual scientists

| | |
|---|---|
| Distinguishing | 1) A reflection of relevant capacities |
| | 2) Sufficiently well correlated with peer review |
| Acknowledging characteristics of the individual career | 3) Independent of career length |
| | 4) A balanced appreciation of collaborative output |
| Fit for common use | 5) Up to date |
| | 6) Easy to calculate |
| Acknowledging the nature of scientific communication in the discipline | 7) Outlier proof |
| | 8) Avoiding bias |
| Resisting human error and interventions | 9) Error proof |
| | 10) Manipulation proof |

*1) A reflection of relevant capacities:* Indicators should reflect the capacities sought in the subjects under evaluation and should be presented as such in the framework of an evaluation. The inclusion of other values without any further comment, merely because they are available, may lead to misperception or even misuse and should be avoided. Also for accepted indicators, a careful reflection of what is captured and what is not, provides important information for a correct interpretation. The most highly cited publications for instance do not always represent a researcher's major scientific breakthroughs [TIJSSEN ET AL., 2002], but may be highly appreciated for other qualities, such as providing a thorough review.

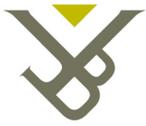

*2) Sufficiently well correlated with peer review:* The best way to judge the performance of an indicator is to compare its results to those of a peer review evaluation that is trusted and well accepted. Experiences have shown that high correlation coefficients between peer review and 'traditional' bibliometric indicators are rarely found [see e.g. RINIA ET AL., 1998; AKSNES AND TAXT, 2004; MOED, 2005]. Own investigations have shown that, besides reliability, comparability of results depends on the nature of the indicators, on the subject area and on the intrinsic characteristics of the methods [RONS AND SPRUYT, 2006].

*3) Independent of career length:* A shorter career should not be a handicap in a selection process for excellent researchers. Excellence programs will indeed seek to fund candidates who will continue to perform at excellence level in the years to come. Bibliometric indicators for individual researchers should therefore account for career length or limit the influence of long past performance on the results.

*4) A balanced appreciation of collaborative output:* Research results and publications are often the product of teamwork. The influence of exceptional output achieved in a large collaboration should be brought to proportion (see also 'Outlier proof'). The indicators should value collaborative outputs, while not overestimating these to a point that the collaborative nature of the output in itself becomes an advantage with respect to others.

*5) Up to date:* Some phenomena, such as extraordinary highly citedness, can only be detected after some time. Thomson ISI identifies Highly Cited Researchers per category based on 20-year rolling time periods, starting with articles and their citations in the years 1981-1999. Such timeframes are not suited to detect researchers who only recently became prominent actors in the field. They also do not guarantee that an identified Highly Cited Researcher is still active, or still working on the same successful research topic. As future performance is a crucial aim of excellence programs, it is important that the indicators used are able to detect a researcher's prominence soon after it is established, expecting it will continue in the years to come.

*6) Easy to calculate:* In the framework of applications for funding, supportive bibliometric information is often desired, yet not included in the files. This leads to a search for indicators that, if not directly available from bibliometric databases, are at least easy to calculate for a substantial amount of applicants. This involves in most cases very basic indicators, like numbers of publications, citations, the h-index or journal impact factors. Yet without a proper interpretation by experts or a proper normalization, these values give biased information on the quality of the work and do not allow a comparison of applications across disciplines. A meaningful indicator designed to support selection procedures needs to combine subtlety with relative ease if it aspires to be used in practice on a regular basis.

*7) Outlier proof:* Certain indicators make averages over elements of possibly very different dimensions. A small number of highly cited articles for instance may substantially influence the mean value, as citation distributions tend to be skewed [SEGLEN, 1992]. Such occasional highly cited, possibly co-authored papers are not necessarily representative for the general performance level, especially for individual scientists, and should not dominate indicators. An indicator limiting the influence of highly cited papers was introduced by HIRSCH [2005] as the h-index. Other techniques focus on the group of highly cited papers [TIJSSEN ET AL., 2002] or consider the distribution of papers over quantiles [ZITT ET AL., 2007].

*8) Avoiding bias:* Indicators may inherently contain bias caused by a variety of choices made in their construction, such as the database used, categories therein, the counting method [GAUFFRIAU AND LARSEN, 2005], normalisation or its absence [ANTONAKIS AND LALIVE, 2008], the observation period or the cut-off values [BASU, 2006]. The extent to which indicators measure performance thus involves, besides the coverage of the publications by the database, many other characteristics of publication and citation behaviour, which become more specific for more narrow topics. Therefore, in the process of constructing indicators for individual scientists, even the smaller choices should be made conscious of the advantage that they may introduce of one researcher over the other.

*9) Error proof:* A critical overview on technical and methodological problems that may arise when gathering and investigating publication and citation data is provided by VAN RAAN [2005]. Citations may for instance be misattributed due to variations and errors in an author's name or in the reference to his/her publication. At higher aggregation levels, several types of errors may be averaged out. In outputs of individual researchers however, they may be very unequally distributed, influencing results to a very large extent. Their effect can be highly diminished when a careful, time consuming screening of publications and citations can be performed. Most selection processes however will require a fast analysis for a limited budget. Indicators designed to quickly assess individual researchers should therefore have a limited sensitivity to data errors.

*10) Manipulation proof:* Not only when using indicators, but already in the process of designing them, attention should go to the possibilities for manipulation by different actors who may be tempted when their situation depends on the outcomes. A number of unintended effects can thus be avoided from the start. The number of citations for instance is harder to manipulate than the number of publications. Because many institutions contribute to it, the effect of certain potential citation-deals remains limited. Another possible strategy is to make it less clear which quantity to optimize to get better results. This is a good reason why evaluations should always use a set of bibliometric indicators and not focus on only one.

**Proposed indicator: the evolution of Impact Vitality**

*The indicator and its characteristics:*

The indicator proposed in this paper represents the profile over time of Impact Vitality (IV). The Impact Vitality measures how the volume of research that is influenced by the scientist's work - represented by the publications that cite it - evolves over time. Citing publications are counted over time with lower weights for higher age, after which a normalization is applied. For the Impact Vitality $IV(y_1, n)$ in year $y_1$, with a window starting $n$ years back in $y_n$, citing publications $P(y_i)$ published in year $y_i$ receive a lower weight when having a higher age $i$:

$$IV(y_1, n) = [n \{ (\Sigma_{i=1 \to n} P(y_i) / i) / \Sigma_{i=1 \to n} P(y_i) \} - 1] / [\{ \Sigma_{i=1 \to n} 1/i \} - 1]$$

with $n > 1$, $y_{i+1} = y_i - 1$ and $\Sigma_{i=1 \to n} P(y_i) > 0$

The normalization is tailored to yield a value larger than 1 when citing publications increase with time, and smaller than 1 when they decrease in time. A value of 1 is produced when the scientist's work receives an equal amount of citations every year. A value of 0 is produced when the last citations received are in the earliest year of the citing publications window. The size *n* of the citing publications window needs to be large enough to bridge the natural yearly fluctuations in the number of citing documents in order to pick up the trend. To illustrate the characteristics of Impact Vitality, Table 2 shows a number of simulated examples. Calculations for a real example can be found in Table 5.

Table 2. Simulated examples illustrating Impact Vitality characteristics for a citing publications window of 5 years

|  |  | Number of citing publications per year |  |  |  |  | Impact Vitality value |
|---|---|---|---|---|---|---|---|
|  |  | $y$ | $y-1$ | $y-2$ | $y-3$ | $y-4$ | $IV(y, 5)$ |
| Case A: | Constant amount of citing documents over time: value = 1 | 5 | 5 | 5 | 5 | 5 | 1.0 |
| Case B: | Increasing amount of citing documents over time: value > 1 | 5 | 4 | 3 | 2 | 1 | 1.5 |
| Case C: | Decreasing amount of citing documents over time: value < 1 | 1 | 2 | 3 | 4 | 5 | 0.5 |
| Case D: | Same proportional increase of citing documents over time (cases B and D): same value | 10 | 8 | 6 | 4 | 2 | 1.5 |
| Case E: | Fluctuating amount of citing documents over time: value according to the final trend, < 1 for a final negative trend | 1 | 2 | 3 | 2 | 1 | 0.8 |
| Case F: | Fluctuating amount of citing documents over time: value according to the final trend, > 1 for final positive trend | 3 | 2 | 1 | 2 | 3 | 1.1 |

The Impact Vitality is multiplier invariant, meaning that multiplying the number of citing publications over the entire period does not affect its value (illustrated by cases B and D in Table 2):

$$IV'(y_1, n) = [n \{ (\Sigma_{i=1 \to n} A\ P(y_i) / i) / \Sigma_{i=1 \to n} A\ P(y_i) \} - 1] / [\{ \Sigma_{i=1 \to n} 1/i \} - 1] = IV(y_1, n)$$

This has the advantage that it treats authors from disciplines with different typical citation rates on an equal basis. Also the relative values for two researchers will be the same in different databases, the coverage of which differs by a factor. With respect to this characteristic, it is important to stress that excellence is not indicated by an Impact Vitality value at one particular moment, but by its profile over time. At one moment in time the same values may indeed be generated for cases where citing publications increase with the same rate, yet at very different absolute levels.

The presented indicator provides an easy to use insight into the way in which the topics treated by the author receive increasing or decreasing attention, thereby highlighting the continuing innovative capacities of a researcher. An Impact Vitality value larger than 1 sustained year after year indicates a continuously growing uptake of the scientist's work in recent developments in the field. For highly influential researchers, this personal impact area may approach the entire set of publications in the topics concerned. The profile over time thus provides a potential tool (in a set of indicators) to detect steadily increasing high performance and impact, which may be used to identify excellent researchers.

*Related indicators:*

The concept of vitality has been used earlier in a bibliometric context by BOYACK AND KLAVANS [2005] for the generation of Maps of Science and applied for instance in thought leadership patterns for national and institutional comparison [KLAVANS AND BOYACK, 2008]. This vitality measure, bound between 0 and 1, is an average of the inverse age of all references from all current papers assigned to a community. A high vitality value is associated with a fast moving research area, where new findings are quickly incorporated in further research efforts.

Sequences in time of bibliometric indicator values were observed earlier by LIANG [2006] with respect to the h-index. Also like the h-index, the Impact Vitality combines publication and citation data, yet in a different way. The h-index counts publications that are cited to a certain extent. The Impact Vitality is based on the publications that cite a body of work.

The Impact Vitality bears structural resemblance to the AR-index introduced by JIN ET AL. [2007], based on the h-index and measuring the h-core's citation intensity. Table 3 shows that, while Impact Vitality and AR-index to some extent share a similar structure, they have a clearly different content.

Table 3. A comparison between AR-index and Impact Vitality

|  | **AR-index** | **Impact Vitality** |
|---|---|---|
|  | $AR = (\sum_{j=1 \to h} cit(j)/age(j))^{(1/2)}$ | $IV = [\{ (\sum_{j=1 \to m} 1/age(j))/(m/n) \} - 1] / [\sum_{i=2 \to n} 1/i]$ |
|  | where $age(j)$ is the age of article $j$, $cit(j)$ is the number of citations to it and $h$ is the number of elements in the h-core [JIN ET AL., 2007] | where $age(j)$ is the age of citing document $j$ and $m$ is the number of citing documents in the time window of $n$ years |
| BASIS: | Numbers of citations | Numbers of citing publications |
| WEIGHTS: | Lower weight for older cited publications | Lower weight for older citing publications |
| LIMITS: | Citations to the h-core | Citing publications in a chosen time frame |
| UNLIMITED: | Age of the citations | Age of the publications cited |
| CONTENT: | Accumulated impact of the most highly cited publications, with an emphasis on the impact of the most recent highly cited publications | Impact of the work in a specified period, with an emphasis on the impact in the most recent years |

*Options and limitations:*

Several kinds of variants of the indicator can be considered. The time window can be a moving window of fixed length, or a window growing with the career from a fixed point. This fixed starting point may for example be the year of PhD or the year in which the earliest citations appear, yielding windows respectively covering the scientist's 'senior' or 'global' career until the year of observation. Other variants could be created by selecting specific cited or citing document types, or by attributing different age-related weights.

The proposed indicator, like any numerical measure, has its limitations. Its multiplier invariance means that it captures patterns rather than volume, so that other indicators will be needed to compare absolute performance levels. While reasonable results can be obtained also for low numbers of citing publications, the indicator should not be used in cases where the number of citing publications fluctuates between zero and non-zero values over the years. Therefore, for research areas where coverage by the publication database is low to this extent, the proposed indicator offers no solution.

**Indicator evaluation**

This section readdresses the requirements for the indicator put forward in the previous sections and discusses to what extent the proposed indicator meets them.

*1) A refection of relevant capacities:* A continuously strong Impact Vitality profile over time reflects a continuously growing uptake of the researcher's work in recent developments in the field. This holds several elements from the definition of excellence derived in the previous sections, focusing on the aspects of continuity and increasing impact in the field. The citing publications counted may partly reflect the continued impact of previous work, possibly on previous topics. Choosing to follow experimental, innovative, high risk research lines, will thus not result in an immediate drop in Impact

Vitality values but might on the contrary result in a noticeable increase in the period of time where the researcher builds up impact in the new field.

*2) Sufficiently well correlated with peer review:* As a first test of the indicator's potential to single out outstanding researchers as recognized by peers, it was applied to a sample of applicants (excluding candidates rejected for formal reasons) to annual open calls for full time senior Research Fellowships aimed at excellent researchers. Each year's call is dedicated to predefined broad research themes. Selection depends on the peer review results and on the number of available Fellowships. Table 4 shows that all selected applicants had senior career Impact Vitality profiles above 1 over the years, and that no applicants were selected with profiles including values below 1 for one or more years. It also shows that the profiles of the selected applicants fluctuate less over a same period in time than those of the non-selected applicants.

Table 4. Peer review based selection vs. Impact Vitality profiles[1]
for VUB Research Fellowship calls 2000-2007[2]

|  | *Selected candidates* | *Not selected candidates* |
|---|---|---|
| Number of candidates | 8 | 17 (2 of which with PhD less than 4 years before call) |
| Number of citing publications per year | 5 years until call: 26 to 399, average 88 | 5 years until call: 10 to 97, average 33 |
|  | PhD until call: 22 to 218, average 57 | PhD until call: 4 to 70, average 24 |
| Minimum $IV_{PhD}$ | 1.05 to 1.76, average 1.27 | 0.36 to 1.64, average 1.09 |
|  | 100% with all $IV_{PhD}$ values > 1 | 60% with all $IV_{PhD}$ values > 1 |
| $IV_{PhD}$ fluctuation[3] in 5 years until call | 0.12 to 0.33, average 0.25 | 0.12 to 2.43, average 0.73 |

[1] Senior career Impact Vitality ($IV_{PhD}$) profiles calculated using the Web of Science for all career years until the Research Fellowship call (November 2008, citing publications windows ≥ 4 years).

[2] Applications in predefined research themes; themes in Social Sciences and Humanities excluded.

[3] (Maximum $IV_{PhD}$ value) - (Minimum $IV_{PhD}$ value), calculated only where 5 $IV_{PhD}$ values are available until the call.

*3) Independent of career length:* Citations to a researcher's older work, in previous phases of a longstanding career, will partly occur together with citations to the more recent work (in recent citing publications) and partly be given less weight (in older citing publications). These mechanisms limit the advantage that a longer career may yield as compared to an indicator based on numbers of citations and disregarding their age. At an early stage of the career, the indicator values are often higher, as the number of citing publications is quickly building up to a normal level for the research domain. A similar effect may take place when a career takes a new turn, starting up a new research topic.

*4) A balanced appreciation of collaborative output:* Collaborative output can be counted in different ways, fractional or whole, while a combination may be the best way to proceed. Such a decision on the counting method is not required for the Impact Vitality measure, as it observes a trend and not an absolute value. Authors from different disciplines will not be advantaged or disadvantaged with respect to each other due to the fact that the typical nature of output in their research area is collaborative or single author.

*5) Up to date:* One may assume that in general, an increase of citing documents (Impact Vitality > 1) will to a large extent concern a scientist's recent work and not a new or revived interest in older ideas. Furthermore, attributing lower weights to older citing publications emphasizes the more recent impact on the field.

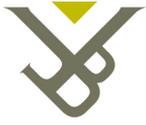

*6) Easily calculated:* A correct selection of the citing publications, based on the cited author's name and initials, is essential. Cited work of homonyms needs to be detected and the according documents excluded. This elimination procedure may be facilitated by the fact that several cited documents by a homonym may be linked to a same citing document. Once these eliminations are completed the calculations are straightforward. The Impact Vitality profile is easily updated yearly, by adding the citing publications from the last year to the data set and calculating the Impact Vitality value for that year.

*7) Outlier proof:* An occasional outlier will modulate the Impact Vitality profile. A paper that is highly cited compared to the others will increase the Impact Vitality value in the first couple of years after its publication. As, in general, citations to a particular publication decrease after a peak in the first years following publication, the Impact Vitality value will in a second phase descend, possibly below the level it would have had without the outlier, and then gradually recover to this level. For outliers that continue to attract a comparable high number of citations yearly, the Impact Vitality value will in the second phase simply descend to the level it would have had without the outlier. The time frame of this process depends on the type of outlier and on the indicator variant (fixed starting point or moving window). The amplitude of the effect will depend on the ratio of the number of documents citing *only* the highly cited paper compared to the number of documents citing other work by the same author, and on whether or not more outliers are present. An illustration of the effect in one real example is included in Table 5.

Table 5. Example of an Impact Vitality profile[1] for a selected author[2]

| | Impact Vitality profile for all WoS-publications citing the author | | Excluding publications citing only the author's most cited paper | | Excluding self-citing publications by the author | |
|---|---|---|---|---|---|---|
| | Citing publications | $IV_{PhD}$ | Citing publications | $IV_{PhD}$ | Citing publications | $IV_{PhD}$ |
| 2007 | 316 | 1.40 | 285 | 1.37 | 307 | 1.40 |
| 2006 | 355 | 1.49 | 314 | 1.44 | 351 | 1.49 |
| 2005 | 341 | 1.52 | 299 | 1.46 | 335 | 1.53 |
| 2004 (call) | 373 | 1.62 | 322 | 1.54 | 363 | 1.62 |
| 2003 | 402 | 1.71 | 344 | 1.61 | 397 | 1.72 |
| 2002 | 398 | 1.76 | 332 | 1.63 | 395 | 1.77 |
| 2001 | 406 | 1.82 | 332 | 1.67 | 402 | 1.83 |
| 2000 | 421 | 1.84 | 346 | 1.70 | 416 | 1.85 |
| 1999[3] | 306 | 1.62 | 291 | 1.59 | 296 | 1.61 |
| 1998 | 211 | 1.42 | 211 | 1.42 | 201 | 1.40 |
| 1997 | 188 | 1.36 | 188 | 1.36 | 183 | 1.36 |
| 1996 | 153 | 1.29 | 153 | 1.29 | 149 | 1.28 |
| 1995 | 164 | 1.32 | 164 | 1.32 | 160 | 1.31 |
| 1994 | 125 | 1.21 | 125 | 1.21 | 121 | 1.20 |
| 1993 | 126 | 1.23 | 126 | 1.23 | 125 | 1.23 |
| 1992 | 120 | 1.20[4] | 120 | 1.20 | 116 | 1.19 |
| 1991 | 87 | 1.04[4] | 87 | 1.04 | 84 | 1.03 |
| 1990 | 77 | | 77 | | 76 | |
| 1989 | 76 | | 76 | | 75 | |
| 1988 (PhD) | 82 | | 82 | | 82 | |
| ... | ... | | | | ... | |

[1] Senior career Impact Vitality ($IV_{PhD}$) profiles calculated using the Web of Science (March 2009, citing publications windows ≥ 4 years).
[2] Selected candidate with the highest number of citing publications per year, from the VUB Research Fellowship calls 2000-2007.
[3] Publication year of the author's most cited publication.
[4] Calculation examples:

$IV_{PhD}$ (1991) = [ 4 { ( 87/1 + 77/2 + 76/3 + 82/4 ) / ( 87 + 77 + 76 + 82 ) } - 1 ] / [ { 1 + 1/2 + 1/3 + 1/4 } - 1 ] = 1.04

$V_{PhD}$ (1992) = [ 5 { ( 120/1 + 87/2 + 77/3 + 76/4 + 82/5 ) / ( 120 + 87 + 77 + 76 + 82 ) } - 1 ] / [ { 1 + 1/2 + 1/3 + 1/4 + 1/5 } - 1 ] = 1.20

*8) Avoiding bias:* By its structure, Impact Vitality avoids several kinds of bias. Firstly, the indicator concerns the *evolution* over time (not an absolute number) of citing *publications* (not citations). It is therefore independent of certain domain characteristics, such as the size of the research community working on a topic and its typical number of citations per publication, facilitating a comparison between disciplines. Furthermore, as no limit is imposed on the age of the cited publications counted, the indicator is able to capture the 'full' impact of a researcher's work, whether the domain has a short or a long typical citation delay. Secondly, the citing publications counted may cite work that itself is *not* included in the database used ('target extended' analysis). This makes the indicator better applicable to disciplines that are not so well covered by the chosen publication database.

*9) Error proof:* Different kinds of errors in lists of references may corrupt a large percentage of the citations to a particular article. An example where this amounts to about 18% is reported by GLÄNZEL ET AL. [2003], with three particular types of errors: an incorrect or missing page number (most frequent), followed by an incorrect publication year and an incorrect first author. As the indicator proposed in this paper is based on publications citing a particular *author*, errors in a reference other than in the author's name will be of no influence. With an error in the cited author's name, the citing publication will not be included in the search results, unless it correctly cites other publications by the same author. Errors in the way that publications of the investigated author's are registered in the database will be of no influence, as the indicator is based on information included in the citing publications.

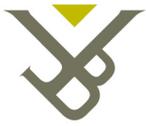

*10) Manipulation proof:* The authors cannot influence each other's or own results by including more than the necessary citations in a same document. Nor does producing more publications that remain uncited influence the indicator value. The publication behavior is not directly steered towards higher impact journals, as this will in itself not influence results, but rather towards sources which can be easily found and accessed by colleagues. Self-citing documents should keep increasing in number over the years in order to help generate Impact Vitality values above 1. The potential amplitude of the effect depends on the ratio of the number of own publications compared to the number of citing publications from other authors. Where judged necessary, self-citing publications can easily be excluded by subtracting them from the total number of citing documents, as is illustrated in Table 5 for a real example.

### Conclusion and further research

A novel indicator is proposed to help identify excellent scientists by referring to a sustained increase of publications that cite their work. It is relatively easy to calculate and hard to manipulate and has a limited sensitivity to outliers in citation counts and to errors in references. Further advantages are a scope broader than the scientist's indexed publications and independence regarding size and citation culture of the research community. While first test results confirm the operational usability, larger scale empirical research is needed before the indicator can be recommended to be included in a set of indicators for particular kinds of assessment of individual scientists. Possible variants leave room for tuning to fit the purposes of different programs. Finally, while designed for the evaluation of individual researchers, applications may also be investigated at other aggregation levels.

\*

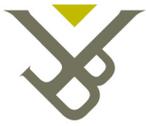